\documentclass[12pt]{article}

\usepackage{amsmath,amssymb,cite,bm}
\usepackage{graphicx}

\makeatletter
 
  \@addtoreset{equation}{section}
 \makeatother

\newcommand{\be}{\begin{equation}}
\newcommand{\ee}{\end{equation}}
\newcommand{\bea}{\begin{eqnarray}}
\newcommand{\eea}{\end{eqnarray}}
\newcommand{\beann}{\begin{eqnarray*}}
\newcommand{\eeann}{\end{eqnarray*}}

\newcommand{\ba}{\begin{array}}
\newcommand{\ea}{\end{array}}

\allowdisplaybreaks

\begin{document}

\setlength{\oddsidemargin}{0cm}
\setlength{\baselineskip}{7mm}

\begin{titlepage}
\renewcommand{\thefootnote}{\fnsymbol{footnote}}
\begin{normalsize}
\begin{flushright}
\begin{tabular}{l}
UTHEP-685
\end{tabular}
\end{flushright}
  \end{normalsize}

~~\\

\vspace*{0cm}
    \begin{Large}
       \begin{center}
         {K\"{a}hler structure in the commutative 
  limit of matrix geometry}
       \end{center}
    \end{Large}
\vspace{0.7cm}

\begin{center}
Goro I{\sc shiki}$^{1),2)}$\footnote
            {
e-mail address : 
ishiki@het.ph.tsukuba.ac.jp},
Takaki M{\sc atsumoto}$^{2)}$\footnote
            {
e-mail address : 
matsumoto@het.ph.tsukuba.ac.jp} and 
Hisayoshi M{\sc uraki}$^{2)}$\footnote
            {
e-mail address : 
hmuraki@het.ph.tsukuba.ac.jp}

\vspace{0.7cm}

     $^{ 1)}$ {\it Center for Integrated Research in Fundamental Science and Engineering (CiRfSE),\\ University of Tsukuba, }
               {\it Tsukuba, Ibaraki 305-8571, Japan}\\
                   
     $^{ 2)}$ {\it Graduate School of Pure and Applied Sciences, University of Tsukuba, }\\
               {\it Tsukuba, Ibaraki 305-8571, Japan}\\

\end{center}

\vspace{0.7cm}

\begin{abstract}
\noindent
We consider the commutative limit of matrix geometry described 
by a large-$N$ sequence of some Hermitian matrices. 
Under some assumptions, we show that the commutative geometry 
possesses a K\"{a}hler structure. We find an explicit relation 
between the K\"{a}hler structure and the matrix configurations 
which define the matrix geometry. 
We also find a relation between the matrix configurations and 
those obtained from the geometric quantization.

\end{abstract}
\vfill

\end{titlepage}
\vfil\eject

\setcounter{footnote}{0}


\section{Introduction}
The matrix geometry is a sort of quantized geometry. It
appears naturally in describing D-branes or membranes in 
string theories and is expected to play important roles 
in formulating the superstring theory and M-theory
\cite{deWit:1988ig,Banks:1996vh,Ishibashi:1996xs}.
The matrix geometry is also closely related to 
interesting fields in mathematics such as the noncommutative 
geometry and the deformation/geometric quantization.

In this paper, we focus on the matrix-geometric 
description of a manifold embedded in the flat $D$-dimensional space 
and understand a relation to the usual differential geometric
description of the manifold.
The matrix geometry for an embedded space in $R^D$ 
can be defined in terms of $D$ Hermitian matrices $X^\mu$.
They can be considered as the quantized version of an embedding 
function $y:{\cal M}\rightarrow R^D$. In fact, in some well-known 
examples such as the fuzzy sphere \cite{Madore:1991bw}, 
these matrices can be constructed explicitly from the 
given embedding function through the geometric quantization.
For example, see \cite{gqtext,gqreview} and references therein for 
this topic.

For matrices with a fixed size, 
one can extract geometric information by using the methods proposed in 
\cite{Berenstein:2012ts} (see also \cite{Schneiderbauer:2016wub}).
Here, the corresponding geometry is defined as loci of zero eigenmodes of 
the matrix version of the Dirac operator, $\gamma_\mu (X^\mu -y^\mu)$ 
(or the Laplacian $(X^\mu -y^\mu)^2$ can also be used  as considered in 
\cite{Schneiderbauer:2016wub}), 
where $y\in R^D$ are parameters.
This definition is natural from the view point of the string theory.
If $X^\mu$ describe a certain shape of a D-brane, the Dirac operator appears 
in the low-energy effective action of
open strings connecting the D-brane with 
another probe brane at the position $y$. 
The zero eigenmodes are just massless modes of 
open strings. Thus at that position, the probe brane feels the
presence of the D-brane. 
So the set of zero eigenstates can naturally be identified with the shape 
of the D-brane.

Here we also remark that 
a similar prescription can also be found in a different context
\cite{Asakawa:2001vm,Terashima:2005ic}. 
Here the Dirac operator corresponds to a tachyon 
configuration in a non-BPS brane system and the set of zero eigenmodes 
corresponds to the shape of the resultant stable D-brane after the tachyon 
condensates.

In this paper, in contrast to the above mentioned work, 
we consider a large-$N$ family of $X^\mu$, not 
just the matrices with a fixed size. 
Namely, our starting point is that $\{X^{(N)}_\mu |N\in I \}$ is 
given, where 
$X^{(N)}_\mu (\mu = 1,\cdots, D)$ are $N\times N$ Hermitian matrices and
$I$ is a fixed index set made of 
a strictly increasing infinite sequence of natural numbers.
Then assuming that the commutative limit is a smooth manifold, 
we try to extract information of the commutative manifold 
from the given matrix configurations. 
In this setup, a description of the commutative space, 
its tangent space and Poisson structure was given in
\cite{Ishiki:2015saa} based on the notion of the coherent states.
We investigate this method further in this paper.

The motivation to consider this setup partially lies 
in the matrix regularization, which
plays an important role in formulating M-theory. 
The matrix regularization is a method of approximating
a functional ring $C({\cal M})$ on a 
symplectic manifold ${\cal M}$ using the matrix algebra $M_N(C)$. 
The regularization is characterized by a sequence $\{T_N \}$ of linear maps 
$T_N:C({\cal M})\rightarrow  M_N(C)$ such that 
in the large-$N$ limit 
the algebraic structure is preserved and 
the Poisson bracket of two functions is mapped 
to the commutator of matrices, namely, 
$||T_N(f)T_N(g)-T_N(fg) || \rightarrow 0$ and 
$||T_N(\{f, g\})-iN[T_N(f),T_N(g)] || \rightarrow 0$.
See \cite{Arnlind:2010ac} for more detailed treatment.
If we regard $X^\mu$ as the image of the embedding function, 
our problem of finding the geometry from given matrices is just 
the inverse problem of the construction of matrix regularization.
Understanding the inverse problem should shed some light on the 
construction problem of the matrix regularization. 




In this paper, for the given sequence of matrices,
we consider how to recover the differential geometry of 
the manifold in the commutative limit.
We focus on matrices which satisfy
\begin{align}
[X^\mu, X^\nu] = \frac{i}{C_N}W^{\mu \nu}(X) + \cdots,
\label{condition}
\end{align}
where, $W^{\mu \nu}(X)$ is a polynomial of $X^\mu$ 
such that its definition (degree and coefficients) does not depend on $N$.
$C_N$ is a real $N$-dependent constant which
goes to infinity as $N\rightarrow \infty$ and $\cdots $ in 
(\ref{condition}) stands for higher order terms in $1/C_N$.
The large-$N$ limit thus corresponds to 
the commutative limit of the matrix geometry.
We also call this limit classical limit in this paper since 
this corresponds to the limit of $\hbar \rightarrow 0$ 
in quantum mechanics, where $X^1 = \hat{p}$ and $X^2 = \hat{q}$. 

We show that if the matrices $X^\mu $ satisfy (\ref{condition})
the associated classical space possesses 
not only a Poisson structure but also a K\"{a}hler structure\footnote{
To conclude this, we also assume one more condition that the ground state 
of the defining Hamiltonian of the classical geometry 
is non-degenerate. This is just a technical assumption
and will be explained in the following sections.
}.
We also argue that when $C_N$ satisfies a certain condition, 
the matrices which satisfy (\ref{condition})
can be semi-classically equivalent to those obtained from the geometric 
quantization scheme.


The condition (\ref{condition}) 
strongly restricts possible form of the matrices.
However, this condition is satisfied by most known symmetric 
examples of matrix geometry such as the fuzzy $CP^n$ 
\cite{Alexanian:2001qj} and fuzzy torus \cite{Connes:1997cr}.
See also \cite{Arnlind:2006ux} 
for a less symmetric example which satisfies
(\ref{condition})\footnote{On the other hand, 
(\ref{condition}) is not satisfied 
for the fuzzy (real) four sphere 
\cite{Castelino:1997rv} for example. }. 
Moreover, even if one adds a perturbation 
given by $N$-independent polynomials in $X^\mu$ to configurations
satisfying (\ref{condition}), 
still the condition is kept satisfied up to
any order of the perturbation. 
Thus, this setup is also relevant for theories 
expanded around some fixed background configuration.

This paper is organized as follows. 
In section \ref{Classical limit of matrix geometry},
we review \cite{Ishiki:2015saa} and see
the description of the classical space and 
its tangent space in terms of given matrix elements.
In section \ref{Kahler structure},
we show that when (\ref{condition}) is satisfied, 
the classical space possesses a K\"{a}hler structure. 
We also relate the geometric structure to the matrix 
configurations.
In section \ref{An example: fuzzy sphere}, 
we consider the fuzzy sphere as an example and 
show explicit forms of the geometric objects defined in the 
previous section.
In section \ref{Relation to the geometric quantization}, 
we discuss a relation to the geometric quantization.
Section \ref{Summary and discussions} is devoted to a summary and 
discussions.

\section{Classical limit of matrix geometry}
\label{Classical limit of matrix geometry}
In this section, we review \cite{Ishiki:2015saa} and recall the 
description of the classical space and its tangent space in terms of 
the given matrix elements.

\subsection{Classical space}
As we described in the previous section, 
we assume that configurations of the coordinate 
matrices, $\{X^{(N)}_\mu |N\in I \}$, is given. 
In the following, we will omit the superscript $(N)$.

We adopt the definition of the classical space proposed in 
\cite{Ishiki:2015saa}. 
We first define an $N\times N$ Hermitian matrix $H$, which we call 
the Hamiltonian, as 
\begin{equation}
H(y) = \frac{1}{2}(X^{\mu}-y^{\mu}{\bf 1})^2 .
\label{hamiltonian}
\end{equation}
where $y^{\mu}(\mu= 1,2,...,D)$ are real parameters.
Then, we define the classical space as the 
loci of zeros of the ground state energy $E_0(y)$ of $H(y)$
in the classical limit $N \rightarrow \infty$.
More specifically, we write
\begin{equation}
{\cal M } = \{ y \in R^D | f(y) =0 \},
\label{def of M}
\end{equation}
where the function $f(y)$ is the large-$N$ limit of the 
ground state energy,
\begin{equation}
f(y) = \lim_{N\rightarrow \infty }E_0(y).
\label{def of f}
\end{equation}

One can understand this definition as follows.
The ground state energy of the Hamiltonian can be written as 
\begin{equation}
E_0(y) = \frac{1}{2}(\Delta X^{\mu})^2 + 
\frac{1}{2}(\langle X^{\mu} \rangle -y^{\mu})^2,
\end{equation}
where $\Delta X^{\mu}$ and $\langle X^{\mu} \rangle$
are the standard deviation and expectation value 
of $X^{\mu}$ with respect to the ground state, respectively.
When $E_0(y)$ is vanishing, both $\Delta X^{\mu}$ and 
$\langle X^{\mu} \rangle -y^{\mu}$ are vanishing. 
Thus, it follows that if the ground state energy 
vanishes, there exists a wave packet which shrinks to that point.
It is easy to show that the converse is also true. Therefore, 
the space ${ \cal M}$ defined by (\ref{def of M}) can be 
regarded as a set of points in $R^D$ such that
there exist shrinking wave functions, which can naturally be identified 
with the classical points\footnote{
Those wave functions correspond to the coherent states in 
quantum mechanics. See \cite{Grosse:1993uq,Grosse:1994ed,Hammou:2001cc}
for applications of coherent states in describing matrix geometries.}.


There is an ambiguity in the definition of the Hamiltonian.
For example, if one takes the square of $H$ as the new Hamiltonian, 
the above arguments about the shrinking wave packets will still hold.
Since the classical geometries described by those different Hamiltonians 
are the same, one may use either of those definitions. 
However, the definition (\ref{hamiltonian}) has two advantages.
One is just the simplicity. 
Our hamiltonian is the simplest choice among those which 
are compatible with the translational and rotational symmetries of $R^D$.
This will make it easier to compute the geometry and the geometric objects
discussed in the next section from the given matrices. 
The other advantage is in the context of the string theory. 
We can find some physical origins of (\ref{hamiltonian}) in some
systems with D-branes. For example, 
we can 
regard (\ref{hamiltonian}) as the 
Laplacian in the low energy effective action of open strings in a
D-brane system with a probe D0-brane 
\cite{Berenstein:2012ts} (see also \cite{Chatzistavrakidis:2011gs}).
The classical space then corresponds to the shape of a D-brane 
(or a membrane) in the target space. 
Thus, the definition (\ref{hamiltonian}) allows
a clear physical interpretation.

For a general (random) sequence of matrices, 
the classical space (\ref{def of M}) is the most likely 
to be a non-manifold or an empty set. 
A necessary condition to have a non-empty set is 
that the matrices $X^\mu$ become commuting with each other in the 
large-$N$ limit, since the ground state energy is bounded from 
below by $| \langle  [X^\mu, X^\nu] \rangle |/2 $, where 
$\mu$ and $\nu$ are arbitrary. This can be shown in the same way as 
the uncertainty inequality in the quantum mechanics.
Since we assume (\ref{condition}) in this paper, this necessary
condition is satisfied. 

The relation (\ref{condition}) also allows us 
to relate the definition of the classical space with that 
in \cite{Berenstein:2012ts}.
In  \cite{Berenstein:2012ts}, 
surfaces embedded in the three dimensional flat space are
considered\footnote{See also \cite{deBadyn:2015sca} for an application to
describing various configurations of membranes and see
\cite{Karczmarek:2015gda} for a higher dimensional extension.}.
Here, the classical space is defined for any fixed $N$ 
as the loci of zero eigenvectors of the Dirac operator 
$D(y)= \sigma_i (X^i-y^i) $, where $\sigma_i$ are the Pauli matrices. 
Note that the square of the Dirac operator gives our Hamiltonian plus 
terms proportional to the commutators of $X^\mu$. 
If the condition (\ref{condition}) is satisfied, the commutators 
become negligible in the large-$N$ limit. Hence, in the large $N$ limit, 
the classical space defined by the Hamiltonian is equivalent to 
the space defined by the Dirac operator.  

This fact allows us to use the method proposed in \cite{Berenstein:2012ts} 
to check whether the configuration defines a smooth geometry or not, at 
least for some particular cases. In this method, 
one considers an index defined from the matrix Dirac
operator. If the index is non-vanishing, it is assured that 
the classical space is a smooth surface embedded in the flat space. 

Though this can be applied in our setup,
we will not consider this problem further in the following.
Instead, 
we assume that the classical space is a smooth manifold such as 
those having non-vanishing values of the index.
More precisely speaking, we assume that ${\cal M}$ is a non-empty 
connected subset of $R^D$ and there exists a neighbourhood of ${\cal M}$ on which the 
function $f(y)$ is smooth. The latter condition is required for 
${\cal M}$ to be smooth and to have a well defined tangent space, as we 
will see in the next subsection.

\subsection{Tangent space}
In order to develop a theory of differential geometry on ${\cal M}$,
here we define the tangent space of ${\cal M}$ in terms of the given matrices.
At each point $y \in {\cal M}$, let us consider 
an arbitrary $D$-dimensional vector $B^{\mu}(y)$. 
There should exist a projection operator, which projects
$B^{\mu}(y)$ onto its tangent components along ${\cal M}$.
The projection operator is given by \cite{Ishiki:2015saa}
\begin{equation}
P^{\mu}{}_{\nu}(y) = \delta^{\mu}_{\nu} - \partial^{\mu}\partial_{\nu}f(y).
\label{Pmunu}
\end{equation}
Here and thereafter, we raise and lower the $D$-dimensional 
indices by the Kronecker delta (the flat metric on the target space), 
so that $\partial^{\mu} = \partial_{\mu}$. 
See appendix \ref{Projectivity} for a proof of the projectivity of 
$P^{\mu}{}_{\nu}$.
In terms of the projection operator, 
the tangent vectors are defined by the relation,
\begin{equation}
P^{\mu}{}_{ \nu}(y)B^{\nu}(y) = B^{\mu}(y),
\label{tangent vectors}
\end{equation}
for $y \in {\cal M}$.

The projection operator is positive definite on the tangent vectors 
by definition and hence it can be regarded as a metric 
on ${\cal M}$.
This metric corresponds to the induced metric 
(the closed string metric \cite{Seiberg:1999vs}) associated with 
the original embedding ${\cal M} \rightarrow R^D$.

The projection $P^{\mu}{}_{\nu}$ can be computed 
from the given matrices by using the perturbation theory
in quantum mechanics, as follows. 
For a sufficiently small shift of the variables, 
$y^{\mu}\rightarrow y^{\mu}+\epsilon^{\mu}$, 
we have $ H(y+\epsilon ) =H(y)+\epsilon \cdot (y-X) + \frac{1}{2}\epsilon^2$. 
The deviation of the ground state energy 
can be computed by treating the terms with 
$\epsilon^{\mu}$ as perturbation.
The large-$N$ limits of the perturbative corrections for the ground state 
energy correspond to the derivatives of $f(y)$, which give
$P^{\mu}{}_{\nu}$.  
More specifically, by introducing the eigenstates of $H(y)$ as
\begin{equation} 
H(y)| n,y\rangle = E_n(y) | n,y \rangle,
\end{equation} 
the projection operator can be written as 
\begin{equation}
P^{\mu}{}_{\nu}(y) = 2\ \lim_{N \rightarrow \infty}
\sum_{n\neq 0}{\rm Re} 
\frac{\langle 0,y | X^{\mu} | n,y \rangle \langle n,y |
X_{\nu}| 0,y \rangle
}{E_n(y)-E_0(y) }.
\label{projection}
\end{equation}
This expression provides a direct relation between the matrix elements 
and the geometric object $\{P^\mu {}_\nu \}$.
Here, in order to
write down the formula (\ref{projection}),
we have assumed that the ground state is non-degenerate 
on ${\cal M}$.
The case with degenerate ground states will be briefly 
discussed in the last section.

For later convenience, we write some useful relations below. 
For any normalized vector $|\psi \rangle$, we have
\begin{align}
\lim_{N \rightarrow \infty} 
\langle
\psi | y^{\mu}-X^{\mu} |0,y 
\rangle =0. \;\;\; (y \in {\cal M})
\label{linear0}
\end{align}
This follows from the Cauchy-Schwarz inequality
and the fact that $\lim_{N \rightarrow \infty}E_0(y)=0$ 
for $y \in {\cal M}$.
This relation can be generalized to 
\begin{align}
\lim_{N \rightarrow \infty} 
\langle
\psi | h(y)-h(X) |0,y 
\rangle =0, \;\;\; (y \in {\cal M})
\label{linear generalized}
\end{align}
where $h(y)$ is any polynomial with the
coefficients and degree $N$-independent.
Here, we have assumed that $X_{\mu}$ are norm-bounded 
in the large-$N$ limit.

\subsection{Local coordinates}
We can introduce at least locally some parameters $\sigma^a $ 
which solve $f(y(\sigma))=0$. The vectors of the form
\begin{equation}
e_a^{\mu}(\sigma) = \frac{\partial y^{\mu}(\sigma)}{\partial \sigma^a}
\label{vielbein}
\end{equation}
are tangent vectors satisfying (\ref{tangent vectors}).
Since we have assumed that ${\cal M}$ is a manifold,
there exists a parametrization such that 
$\sigma^a$ are local coordinates on ${\cal M}$, where 
$a$ runs from $1$ to the dimension of ${\cal M}$.
In this case, the vectors (\ref{vielbein})
form a basis of the tangent vectors at $y(\sigma )\in {\cal M}$
and the symmetric tensor 
\begin{equation}
g_{ab}(\sigma) = (e_a(\sigma), e_b(\sigma)) 
= \frac{\partial y^{\mu}(\sigma)}{\partial \sigma^a}
\frac{\partial y_{\mu}(\sigma)}{\partial \sigma^b}
\label{induced metric}
\end{equation}
is non-degenerate, where $(\; , \;)$ is the natural inner 
product of the $D$-dimensional vectors in $R^D$.

Any tangent vector $B^{\mu}$ 
can then be expanded using (\ref{vielbein}) as 
\begin{equation}
B^{\mu}= B^{a}e^{\mu}_a.
\end{equation}
The coefficients $B^a$ are uniquely determined from $B^{\mu}$, once 
the basis is fixed.
Similarly, for a given differential form on $R^D$, 
we can also consider a pullback onto ${\cal M}$. 
For example, a one-form $B = B_{\mu}dy^{\mu}$ gives 
a one-form $B_a d\sigma^a$ on ${\cal M}$, where 
$B_a = B_{\mu}e^{\mu}_a$.
Thus by using the local coordinates, 
we can always move to the usual convention that the 
indices of differential forms and tangent vectors run from 
1 to the dimension of the manifold under consideration.
For example, (\ref{induced metric}) is nothing but 
the induced metric $\delta_{\mu \rho }P^{\rho}{}_{\nu}$
written in terms of the local coordinates.

In this way, switching the conventions of indices is always possible, once 
local coordinates are introduced.
However, we will mainly work with the indices $\mu, \nu$ in this paper
to avoid complexity of introducing the local coordinates.

\section{K\"{a}hler structure}
\label{Kahler structure}
In this section, we show that when the condition 
(\ref{condition}) is satisfied, 
the classical space ${\cal M}$ possesses a K\"{a}hler structure, which is 
a compatible triple of Riemann, symplectic and complex structures.

\subsection{Poisson structure}
Before we consider the K\"{a}hler structure, we introduce
a Poisson structure, which becomes very important in this section.
Let us consider a $D\times D$ real antisymmetric 
matrix defined as\footnote{
Note that we 
use the same notation for this matrix and the polynomial in 
the right-hand side of (\ref{condition}). 
This is because the matrix (\ref{Poisson})
is indeed obtained by replacing $X^\mu $ with  $y^\mu $ in the polynomial 
in (\ref{condition}). See (\ref{linear generalized}).
}
\begin{equation}
W^{\mu \nu}(y) = -i \lim_{N \rightarrow \infty}
C_N \langle 0,y | [X^{\mu}, X^{\nu}] | 0,y \rangle.
\label{Poisson}
\end{equation}
For $y\in {\cal M}$, 
one can see that (\ref{Poisson}) gives a Poisson tensor on 
${\cal M}$ \cite{Ishiki:2015saa}.
Namely, it is a tangent bivector on ${\cal M}$ satisfying
\begin{equation}
P^{\mu}{}_{\nu}(y)W^{\nu \rho}(y)=W^{\mu \rho}(y), \;\;\; (y \in {\cal M})
\label{W tangent}
\end{equation} 
and the Poisson bracket
defined by 
\begin{equation}
\{f, g \}=W^{\mu \nu}(\partial_{\mu}f)(\partial_{\nu}g)
\label{P bracket}
\end{equation}
satisfies the Jacobi identity. 
See appendix \ref{Properties of W} for a proof of the above 
statements.

\subsection{Symplectic structure}
\label{sec sym str}
Since our Hamiltonian depends on the parameters $y^{\mu}$,
we can introduce the notion of the Berry phase and the Berry curvature. 
Here, we show that under our assumptions, the Berry curvature of the 
Hamiltonian (\ref{hamiltonian})
gives a symplectic form on ${\cal M}$.

Under an infinitesimal shift of the parameters, 
$y^{\mu} \rightarrow y^{\mu}+ \epsilon^{\mu}$,
the ground state of the Hamiltonian (\ref{hamiltonian})
varies by $|0,y \rangle \rightarrow |0,y \rangle + 
\epsilon^{\mu}\partial_{\mu}|0,y \rangle$, where 
$\partial_{\mu}$ here means a linear map on the Hilbert space 
defined by (\ref{derivative of gs}), which 
is just the first order formula of the perturbation theory.
$A_{\mu}(y) \in R$ in (\ref{derivative of gs}) 
is a real c-number and called the Berry connection.
By a phase rotation of the ground state, 
$A_{\mu}(y)$ transforms as a gauge field.
Though at least locally one can eliminate $A_{\mu}(y)$ by a 
suitable gauge transformation, the curvature 
is gauge invariant and is nonvanishing in general. 

It is appropriate to define the covariant derivative on 
the Hilbert space as 
\begin{align}
D_{\mu}= \partial_{\mu}-iA_{\mu}.
\end{align}
Then the Berry curvature is defined as the 
field strength of $A_{\mu}(y)$:
\begin{align}
F_{\mu \nu}=i[D_{\mu}, D_{\nu}].
\label{Berry curvature}
\end{align}
$F_{\mu \nu}$ is a real number and can also be written as 
\begin{align}
F_{\mu \nu}(y)= 2
\sum_{n\neq 0}
{\rm Im} 
\frac{\langle 0,y|X_{\mu}|n,y \rangle  \langle n,y|X_{\nu}|0,y \rangle
}{(E_n(y)-E_0(y))^2}.
\end{align}
Note that $F_{\mu \nu}$ is defined for each fixed $N$ unlike 
$P^{\mu}{}_{\nu}$ and $W^{\mu\nu}$, which are defined only 
in the large-$N$ limit.

We define a 2-form on $R^D$ by 
\begin{align}
\omega_{\mu \nu}(y) = -\lim_{N\rightarrow \infty} \frac{F_{\mu \nu}(y)}{C_N}
=- \lim_{N \rightarrow \infty }
\frac{2}{C_N}
\sum_{n\neq 0}
{\rm Im} 
\frac{\langle 0,y|X_{\mu}|n,y \rangle  \langle n,y|X_{\nu}|0,y \rangle
}{(E_n(y)-E_0(y))^2}.
\label{symplectic}
\end{align}
The pullback of this 2-form gives a symplectic form on ${\cal M}$ 
as we will show below.
Firstly, it satisfies
\begin{align}
W^{\mu \rho}(y)\omega_{\rho \nu}(y)= P^{\mu}{}_{\nu}(y).
\;\;\; (y \in {\cal M})
\label{symp 1}
\end{align}
Namely, on tangent vectors it gives the inverse of the Poisson tensor.
This guarantees that the pullback is non-vanishing everywhere on 
${\cal M}$.
Secondly, it satisfies $d\omega =0$ or equivalently, 
\begin{align}
\partial_{\mu}\omega_{\nu \rho }(y) + ({\rm cyclic \; permutation})=0.
\;\;\; (y \in {\cal M})
\label{symp 2}
\end{align}
Since taking the pullback and exterior derivative commute,
this implies that the pullback is closed.
The proof of (\ref{symp 1}) is almost the same as that for 
(\ref{W tangent}) shown in appendix
\ref{Properties of W}, while
(\ref{symp 2}) follows immediately from the definition 
of the Berry curvature (\ref{Berry curvature}).
From these properties we find that the pullback of 
$\omega$ gives a symplectic form on ${\cal M}$.

The Berry connection has a physical interpretation in string theories.
By investigating a simple example like a fuzzy sphere plus 
fluctuations, we can find that the Berry connection
corresponds to the gauge fields on D-branes.
Let us consider Myer's effect \cite{Myers:1999ps} and 
regard the fuzzy sphere as the 
blowing up effect from D0-branes to spherical D2-branes.
The fluctuations on D0-branes can be decomposed to 
the tangent and normal components on the sphere\footnote{See 
\cite{Ishiki:2006yr} for example.}.
In our setup, through a direct computation, we can 
see that the tangent components give a fluctuation of 
the Berry connection $A_{\mu}$. 
On the other hand, from the action of D0-branes, we can see that 
the tangent components become the gauge fields on the D2-branes
\cite{Maldacena:2002rb,Ishii:2008ib}.
This suggests that $A_{\mu}$ can be identified with the 
gauge fields on the emergent D-branes.

From (\ref{symplectic}) and (\ref{symp 1}), 
we see that the gauge fields have 
a field strength $F_{\mu \nu}$ with magnetic flux of order $C_N$, which
goes to infinity in the large-$N$ limit.
Since the presence of 
magnetic flux is equivalent to a nontrivial B-field background 
in the Dirac-Born-Infeld action, 
the large-$N$ commutative limit
corresponds to a limit of a very strong B-field background in 
this context.

\subsection{Complex structure}
For a given symplectic structure and an arbitrary metric,
one can construct an almost complex structure, which is 
compatible with the symplectic structure, as shown in appendix 
\ref{Construction of compatible structures}.

Since now we have the symplectic structure 
(\ref{symplectic}) and 
the induced metric (\ref{induced metric}) on ${\cal M}$, 
we can also obtain a compatible almost complex structure 
on ${\cal M}$. It is given by 
\begin{align}
J(y)= 
\frac{1}{\sqrt{W(y)W^T(y)}}
W(y).
\label{complex structure}
\end{align}
Here, $W$ denotes a 
$D\times D$ antisymmetric real matrix with 
the matrix elements given by 
$W^{\mu}{}_{\nu}=\delta_{\nu\rho}W^{\mu \rho}$ and 
the product in (\ref{complex structure}) is just the matrix product.
The inverse power of $W$, $\frac{1}{\sqrt{WW^T}}$, is defined as follows.
In general, from (\ref{W tangent}), $W$ has 
vanishing eigenvalues on the space of normal vectors on ${\cal M}$ 
(the vectors which vanish under a multiplication of 
$P^{\mu}{}_{\nu}$.). 
The inverse of $\frac{1}{\sqrt{WW^T}}$ is taken only in the 
space of the tangent vectors on which $W$ is non-degenerate 
because of (\ref{symp 1}), leaving the zero eigenvalues as they are.

More specifically, for each $y\in {\cal M}$, 
one can take a suitable basis such that $W$ becomes 
the canonical form which consists of some $2\times 2$ diagonal blocks 
as well as some zero eigenvalues.
Denoting the transition matrix to such basis by $V$, we can write
\begin{align}
& \{P^{\mu }{}_{\nu}\}
=
V\left(
\begin{array}{cc}
0 &  \\
 & 1 \\
\end{array}
\right) V^{-1},
\nonumber\\
&
\{ W^{\mu}{}_{\nu} \} = 
V \left(
\begin{array}{cc}
0 &  \\
 & \tilde{W} \\
\end{array}
\right)V^{-1},
\label{diagonalization}
\end{align}
where the up-left and the bottom-right blocks correspond to 
the spaces of the normal and the tangent vectors with dimensions
$D- {\rm dim}{\cal M}$ and ${\rm dim}{\cal M}$, respectively.
Because of (\ref{symp 1}), 
$\tilde{W}$ is a nondegenerate real antisymmetric matrix, which
consists of $2\times 2$ diagonal blocks.
Then, $\frac{1}{\sqrt{WW^T}}$ is defined by
\begin{align}
\frac{1}{\sqrt{WW^T}}
=
V \left(
\begin{array}{cc}
0 &  \\
 & \frac{1}{\sqrt{\tilde{W}\tilde{W}^T}} \\
\end{array}
\right)V^{-1}.
\label{1oversqrtwwt}
\end{align}

One can easily check that (\ref{complex structure}) satisfies
\begin{align}
P(y) J(y)=J(y)P(y)=J(y), \;\; \; J^2(y) = -P(y).  \;\;\;
(y \in {\cal M})
\label{property of J}
\end{align}
The first equation guarantees that $J$ is a linear map from
tangent vectors to themselves and the second equation says that 
$J^2$ is equal to the minus of the identity on the tangent space.
Thus, $J$ defines an almost complex structure on ${\cal M}$.

Furthermore, we can show that $J$ is integrable, so it gives 
a complex structure on ${\cal M}$. 
To show this, let us consider differentiating the 
first equation in (\ref{diagonalization}). We obtain
\begin{align}
\partial_\rho P^\mu {}_{\nu}
+[\Gamma_{\rho }, P]^\mu {}_{\nu}= 0, 
\label{nablaPeq0}
\end{align}
where the derivative shall be restricted to tangential 
directions on ${\cal M}$ and
\begin{align}
\Gamma_{\rho }= V \partial_\rho V^{-1}.
\end{align}
This expression motivates us to define a covariant derivative as,
\begin{align}
\nabla_\rho B^\mu = \partial_{\rho}B^\mu + (\Gamma_{\rho})^\mu {}_{\nu}B^\nu,
\;\;\;\;
\nabla_\rho B_\mu = \partial_{\rho}B_\mu - (\Gamma_{\rho})^\nu {}_{\mu}B_\nu.
\end{align}
Then (\ref{nablaPeq0}) is written as $\nabla P=0 $. 
Since the tangent space of ${\cal M}$ is characterized by $P^\mu{}_\nu $ 
as in (\ref{tangent vectors}), 
the equation $\nabla P=0 $ 
guarantees that $\nabla$ is a linear map from tangent vectors to themselves.
So it is indeed a covariant derivative on ${\cal M}$ preserving 
inner products with respect to the induced metric.
Now, let us consider the covariant derivative of the almost complex 
structure (\ref{complex structure}). 
From (\ref{1oversqrtwwt}) and 
the second equation in (\ref{diagonalization}), 
it is obvious that 
\begin{align}
\nabla_{\rho} J^{\mu}{}_\nu = 0.
\end{align}
Namely, under any parallel transport defined by $\nabla$,
the reference of the holomorphicity set by $J^{\mu}{}_\nu$ 
is preserved. This is nothing but the integrability of $J^\mu{}_\nu$.

The complex structure $J$ can also be written in terms of 
the matrix elements as 
\begin{align}
J^{\mu}{}_{\nu}= 
2\lim_{N\rightarrow \infty }
\sum_{n\neq 0}
{\rm Im} 
\frac{\langle 0,y|X^{\mu}|n,y \rangle  \langle n,y|X_{\nu}|0,y \rangle
}{E_n(y)-E_0(y)}.
\label{ex form of J}
\end{align}
This is proved in appendix 
\ref{Explicit form of J and g}.

\subsection{Riemann structure}
For a given compatible pair of a symplectic form and 
an almost complex structure $(\omega, J)$, one can find a compatible 
Riemannian metric $\tilde{g}$. See appendix 
\ref{Construction of compatible structures}.
In our case, the metric is given by 
\begin{align}
\tilde{g}= \frac{1}{\sqrt{WW^T}}.
\label{compatible metric}
\end{align}
Since $J$ is integrable, 
the compatible triple $(\omega, J, \tilde{g})$ gives 
a K\"{a}hler structure of ${\cal M}$.

The compatible metric $\tilde{g}$ can also be written in terms of 
the given matrix elements as 
\begin{align}
\tilde{g}_{\mu \nu}=
 \lim_{N \rightarrow \infty}
\frac{2}{C_N}
\sum_{n\neq 0}{\rm Re} 
\frac{\langle 0,y | X_{\mu} | n,y \rangle \langle n,y |
X_{\nu}| 0,y \rangle
}{(E_n(y)-E_0(y))^2 }.
\label{ex form of gtilde}
\end{align}
This is shown in appendix
\ref{Explicit form of J and g}.

The compatible metric can be interpreted as the information metric 
of the ground states $\{|0,y \rangle | y\in {\cal M}\}$.
In general, if states are labelled by some parameters 
$\{\lambda^A | A=1,2,\cdots \}$
as $|\lambda \rangle$, 
the information metric (the Bures distance) between those states are defined by
$1-|\langle \lambda |  \lambda' \rangle |^2$. 
In our case, by putting $|  \lambda \rangle = |0,y \rangle $ and 
$|  \lambda' \rangle = |0,y+dy \rangle $, 
we can see that the information metric 
takes exactly the same form as (\ref{ex form of gtilde}).

The metric (\ref{compatible metric}) also has a physical interpretation.
As discussed in section \ref{sec sym str}, the symplectic 
form can be identified with the B-field on D-branes up to an overall constant
and the large-$N$ limit corresponds to a very strong B-field background.
The square of the compatible metric $\tilde{g}$ then takes 
the form $(\tilde{g}^2)_{\mu\nu} 
\propto B_{\mu \rho}\delta^{\rho \sigma}B_{\sigma \nu}$. 
This is nothing but the open string metric in the strong B-field limit 
discussed in \cite{Seiberg:1999vs}.

\section{An example: fuzzy sphere}
\label{An example: fuzzy sphere}
In this section, we consider the fuzzy sphere and give explicit 
forms of the geometric objects defined in the previous section. 

The fuzzy sphere is defined by 
\begin{align}
X^\mu =  \frac{2}{\sqrt{N^2-1}}L^\mu, 
\end{align}
where $L^\mu$ are the $N$-dimensional irreducible representation 
matrices of the $SU(2)$ generators.
The normalization factor is just chosen so that the sphere has 
the unit radius, namely, $(X^\mu)^2 =  1$.
The matrices satisfy the commutation relation, 
\begin{align}
[X^\mu, X^\nu ]= \epsilon^{\mu \nu \rho}\frac{2i}{\sqrt{N^2-1}}X_\rho
=\epsilon^{\mu \nu \rho}\frac{2i}{N}X_\rho + {\cal O}(1/N^3).
\end{align}
Hence the condition (\ref{condition}) is satisfied, where $C_N =N$.

The ground state energy of the Hamiltonian 
is given by 
\begin{align}
E_0(y)=\frac{1}{2}(1+|y|^2) -\frac{N}{\sqrt{N^2-1}}|y|
\end{align}
where $|y|=\sqrt{y_\mu y^\mu}$.
The function (\ref{def of f}) is given by $f(y)=\frac{1}{2}(1-|y|)^2$,
so that the classical space (\ref{def of M}) is $S^2$ with the 
unit radius.
The ground states are given by the so-called Bloch coherent states. 
See \cite{Ishiki:2015saa} for a detailed description.

By using the definition of the coherent states and the expressions 
such as (\ref{projection}), 
we can easily derive explicit forms of 
the geometric objects. They are given as follows.
\begin{align}
&P^\mu {}_{\nu}(y)= \frac{1}{|y|}
\left(\delta^{\mu}_\nu- 
\frac{y^\mu y_\nu }{|y|^2}\right),
\nonumber\\
&W^{\mu \nu}(y) = 2\epsilon^{\mu \nu \rho }\frac{y_\rho }{|y|},
\nonumber\\
&\omega_{\mu \nu }(y)= 
-\frac{1}{2}\epsilon_{\mu\nu\rho } \frac{y^\rho }{|y|^3},
\nonumber\\
&J^\mu {}_{\nu}(y)= \epsilon^{\mu \nu \rho }\frac{y_\rho }{|y|^2},
\nonumber\\
&\tilde{g}_{\mu \nu}(y)= \frac{1}{2|y|^2}\left(\delta_{\mu\nu}- 
\frac{y_\mu y_\nu }{|y|^2}\right).
\end{align}
We can see that for $y \in {\cal M}$ (i.e. for $|y|=1$), they satisfy the required properties such as (\ref{symp 1}), (\ref{property of J}) and so on. 


\section{Relation to the geometric quantization}
\label{Relation to the geometric quantization}
In this section, we show that when $C_N$ in (\ref{condition}) 
satisfies a certain condition, the matrix sequence we have considered 
can be semi-classically equivalent to a sequence obtained through 
the geometric quantization of the K\"{a}hler manifold.
In appendix \ref{Geometric quantization on Kahler manifold}, 
we review the geometric quantization scheme for 
a compact K\"{a}hler manifold.
The main objects in the geometric quantization are the 
holomorphic sections of a complex line bundle, which corresponds 
to wave functions of the quantum mechanics, and 
physical observables represented as some differential operators on 
the wave functions.

We first show that in the large-$N$ limit 
the ground states of the Hamiltonian behave
as holomorphic sections, where the connection of the line bundle is 
given by the Berry connection.
Let us consider an arbitrary antiholomorphic vector field $B^\mu$ satisfying 
\begin{align}
J^{\mu}{}_{\nu}(y)B^{\nu}(y)=
-iB^{\mu}(y). \;\;\; (y \in {\cal M})
\end{align}
The covariant derivative of $\langle \psi | 0,y \rangle $ 
along this vector field, where $ \langle \psi |$ is any normalized state 
independent of $y$, is vanishing in the large-$N$ limit as we will see 
below. Firstly, from the definition of the covariant derivative, we have
\begin{align}
D_{\mu}\langle \psi | 0,y \rangle 
= \sum_{n\neq 0} \frac{ \langle \psi |n,y \rangle 
\langle n,y | X_\mu | 0,y \rangle}{E_n(y)-E_0(y)}.
\label{Dpsi0}
\end{align}
Here recall that as one can see from (\ref{useful}),
multiplying the Poisson tensor on
$\langle n,y | X^\mu | 0,y \rangle$ gives a factor of
$-iC_N (E_n(y)-E_0(y))$ for $y\in {\cal M}$ in the large-$N$ limit.
Thus, we obtain
\begin{align}
J^{\mu}{}_{\nu}(y)D_{\mu}\langle \psi | 0,y \rangle 
&\sim i 
D_{\nu}\langle \psi | 0,y \rangle,
\label{hol section}
\end{align}
where $\sim $ stands for an equality for leading-order terms in 
the large-$N$ limit.
By using this relation, we find that
\begin{align}
B^\mu D_{\mu}\langle \psi | 0,y \rangle 
=i B^\nu J^{\mu}{}_{\nu} D_{\mu}
\langle \psi | 0,y \rangle 
\sim 
- B^\nu D_{\nu}\langle \psi | 0,y \rangle 
\sim 0.
\label{mechanism hol}
\end{align}
Thus in the large-$N$ limit, $\langle \psi | 0,y \rangle $ 
behaves as a holomorphic section.

Secondly, we show that matrix elements of $X^\mu $
can be represented as the form of operators in the geometric quantization,
which act on the holomorphic sections. 
We use the following relation:
\begin{align}
\frac{i}{C_N}W^{\mu \nu}(y)D_{\nu}\langle \psi | 0,y \rangle 
&\sim \sum_{n\neq 0}{ \langle \psi |n,y \rangle 
\langle n,y | X^\mu | 0,y \rangle}
\nonumber\\
&\sim  \langle \psi |X^\mu | 0,y \rangle -  
 \langle \psi |0,y \rangle 
\langle 0,y |X^\mu | 0,y \rangle.
\label{WDpsi0}
\end{align}
This relation (\ref{WDpsi0}) can be rewritten into 
\begin{align}
\langle \psi |X^\mu | 0,y \rangle
\sim
\left(
y^\mu+ \frac{i}{C_N}W^{\mu \nu}(y)D_\mu 
\right)\langle \psi |0,y\rangle, 
\label{operator form}
\end{align}
where we have used (\ref{linear0}).
This form is equal to the prequantum operator 
for the embedding function $y:{\cal M}\rightarrow R^D$ 
obtained by the geometric quantization.
Furthermore, by multiplying $B^\mu D_{\mu}$ to the 
above equation, where $B^\mu$ is any antiholomorphic
vector field, we see that the left-hand side is vanishing
as shown in (\ref{mechanism hol}).  
Thus, we find that the right-hand side 
is also vanishing under the action of $B^\mu D_{\mu}$
at least in the large-$N$ limit. 
This shows that the action of the prequantum operator is 
closed in the space of the holomorphic sections 
of the form $\langle \psi | 0,y \rangle $ in the 
large-$N$ limit. 

These structures are very similar to the construction in 
the geometric quantization.
However, we should note the following point. 
In our setup, the monopole charge of the Berry connection is of order of $C_N$ 
as we discussed in section \ref{sec sym str}.
Let us write $C_N =c N^a$ 
so that the constant 
$a$ represents the order of the monopole charge.
As shown in appendix
\ref{Geometric quantization on Kahler manifold},
the dimension of the Hilbert space spanned by all square integrable 
holomorphic sections is then of 
${\cal O}(N^{a({\rm dim}_{C}{\cal M})})$ in 
the large-$N$ limit, where ${\rm dim}_{C}{\cal M}$ is the 
complex dimension of ${\cal M}$.
On the other hand, the dimension of the space spanned by 
the holomorphic sections of the form $\langle \psi |0,y \rangle $
is at most $N$. Thus, we generally have 
\begin{align}
1/{\rm dim}_{C}{\cal M} \leq a .
\label{inequality}
\end{align}
The equality can be satisfied if any holomorphic section can be written in 
the form $\langle \psi |0,y \rangle $ and 
the ground states $\{|0,y \rangle | y\in{\cal M} \}$ 
span the entire $N$-dimensional vector space. 

If the equality holds in (\ref{inequality}), 
the matrix elements (\ref{operator form}) 
can be a true operator over the entire Hilbert space of the 
holomorphic sections. 
In this case, $X^\mu $ are 
semi-classically equivalent to the operators obtained 
from the geometric quantization.
This case includes well-known examples such as 
the fuzzy torus and fuzzy sphere. 
On the other hand, if the equality is not satisfied, 
the ground states of the Hamiltonian 
are not sufficient to span the whole quantum 
Hilbert space in the geometric quantization.


\section{Summary and discussions}
\label{Summary and discussions}
In this paper, we considered the classical (commutative) limit of 
matrix geometry described by a sequence of 
$D$ Hermitian matrices, $\{(X^{(N)}_1, \cdots, X^{(N)}_D) |N \in I \}$.
We adopted the formulation in \cite{Ishiki:2015saa} 
to describe the classical space and its tangent space. 
The classical space ${\cal M}$ is defined in terms of 
the Hamiltonian (\ref{hamiltonian}) 
and is given by the loci of zeros of the 
ground state energy in the large-$N$ limit as (\ref{def of M}), while
the tangent space of ${\cal M}$ is characterized by 
the projection matrix (\ref{Pmunu}).
We focused on the case where ${\cal M}$ is a smooth manifold and 
the matrices satisfy (\ref{condition}). 
We showed that in this case the classical space possesses a 
K\"{a}hler structure.
Under the assumption that the ground state energy is nondegenerate 
on ${\cal M}$, we found formulas
(\ref{projection}), (\ref{symplectic}), (\ref{ex form of J}) and 
(\ref{ex form of gtilde}), which explicitly relate
geometric structures and the matrix configurations.
We also argued that when (\ref{condition}) is satisfied and 
the constant $C_N$ satisfies the equality in 
(\ref{inequality}), 
the matrix configurations can be semi-classically equivalent to 
those obtained from the geometric quantization.


The inequality (\ref{inequality}) 
shows a lower bound of the order of 
$C_N$ in the large-$N$ limit. We expect that there must also be 
an upper bound beyond which one of our assumptions is no longer satisfied. 
For example, consider the case where $a$ in (\ref{inequality}) is infinity,
this corresponds to situations where
the matrices $X^\mu$ are commuting with each other even at finite $N$.
In this case, the classical space is just a discrete set of points made of $N$ 
eigenvalues of the matrices.
Thus, we do not have the notion of the tangent space etc.
The upper bound of $a$ within which the classical space forms 
a smooth manifold should also be clarified, 
though this is beyond the scope of this paper.

In this paper, 
we have considered only the case where the ground state of the 
Hamiltonian is nondegenerate on the classical space ${\cal M}$. 
Though we think this should be a technical assumption and able to be 
removed without any considerable change of the method, 
there is one thing which we should take into account in the degenerate case.
If ground states are degenerate, the Berry phase is allowed to take 
values in the Lie algebra of the special unitary group with a higher rank.
In this case, we need to deal with the non-abelian gauge fields, as is 
expected from the viewpoint of D-branes in string theory.
It would be interesting to investigate the matrix geometry in this case 
to gain a deeper understanding of multiple D-brane system in string theory.

The geometric objects defined in this paper provide a class of observables
in matrix models, since they are invariant under the 
gauge transformations $X^\mu \rightarrow UX^\mu U^\dagger$. 
These observables capture geometric information of matrix models.
For example, it is an interesting problem to consider a perturbation around 
some fixed matrix configuration and see how the geometric structures 
in matrix models are affected by the perturbation. 

The perturbative calculation of the geometric structures
is able to be carried out in our setting in the large-$N$ limit.
In order to calculate them without any perturbation or extrapolation
of the matrix size, however,
generalization of our work to the case with 
a finite fixed matrix size seems to be needed 
(Recently, this has been partially done in \cite{Schneiderbauer:2016wub}.).
This would make it possible to directly compute the geometric objects 
in matrix models by performing Monte Carlo simulations 
for example \cite{Catterall:2007fp,Anagnostopoulos:2007fw}.

There are also various interesting attempts to understand how the 
matrix models describe gravity \cite{Hanada:2005vr,Steinacker:2007dq,
Steinacker:2012ct,Ito:2015mxa,Hanada:2016zxj}.
It is important to consider whether and how our results can be applied 
in these contexts. 


We hope to report on these issues in the near future.

\section*{Acknowledgments}
This work was supported, in part, by Program to Disseminate Tenure Tracking 
System, MEXT, Japan.

\appendix

\section{Projectivity of $P^{\mu}{}_{\nu}$}
\label{Projectivity}
In this Appendix, we show that $P^{\mu}{}_{\nu}$ 
defined in (\ref{Pmunu}) is a projection from vectors on 
$R^D$ to tangent vectors on ${\cal M}$.

Firstly, we write
$P^{\mu}{}_{\nu}(y)$ at $y \in {\cal M}$ as 
\begin{align}
P^{\mu}{}_{\nu}(y)
= \delta^{\mu}_{\nu} - 
\lim_{N \rightarrow \infty} 
\partial_{\nu}
\langle 0,y| y^\mu-X^\mu |0,y \rangle .
\end{align}
Since $\langle 0,y| y^\mu-X^\mu |0,y \rangle$ is vanishing 
for any $y \in {\cal M}$ in the large-$N$ limit 
(see (\ref{linear0})),
its derivatives along tangent directions
are also vanishing. Thus, for a tangent vector 
$\epsilon^{\mu}_\parallel$ at $y \in {\cal M}$, we obtain
\begin{align}
P^{\mu}{}_{\nu}(y)\epsilon^{\nu}_\parallel
=\epsilon^{\mu}_\parallel.
\label{projectivity1}
\end{align}
This shows that $P^{\mu}{}_{\nu}$ is closed on tangent vectors 
and is equal to $\delta^{\mu}_{\nu}$.

Then, let $\epsilon^{\mu}_\perp$ be a normal vector at 
$y \in {\cal M}$. Below, we will show that 
\begin{align}
P^{\mu}{}_{\nu}(y)\epsilon_{ \perp \mu } \epsilon^{\nu}_\perp
=0.
\label{projectivity2}
\end{align}
From (\ref{projectivity1}) and (\ref{projectivity2}),
the projectivity of $P^{\mu}{}_{\nu}(y)$ 
follows immediately.
Note that the left-hand side of 
(\ref{projectivity2}) is obtained by computing $f(y + \epsilon_{\perp})$.
Namely, when $f(y + \epsilon_{\perp})$ 
is expanded in a power series in $\epsilon_{\perp}$, 
the leading term is equal to
$\frac{1}{2}(\delta^{\mu}_{\nu}- P^{\mu}{}_{\nu}(y))\epsilon_{ \perp \mu } \epsilon^{\nu}_\perp$, which contains the left-hand side of (\ref{projectivity2}).
On the other hand,  $f(y + \epsilon_{\perp})$ can also be evaluated as
\begin{align}
f(y + \epsilon_{\perp}) = \frac{1}{2}|\epsilon_\perp|^2 + {\cal O}(|\epsilon_\perp|^3).
\label{projectivity2prime}
\end{align}
This leads to (\ref{projectivity2}).
In the following, we prove (\ref{projectivity2prime})

In order to prove (\ref{projectivity2prime}), we first prove
\begin{align}
\lim_{N\rightarrow \infty} 
K_N \langle 0, y_1 | H(y) | 0, y_2 \rangle
= \frac{1}{2}(y^{\mu}_2 -y^{\mu})^2
\delta^{(D)} (y_1-y_2), \;\;\; (y_1, y_2 \in {\cal M})
\label{lemma}
\end{align}
where $K_N$ is a positive constant defined below.
The object on the left-hand side can be computed as
\begin{align}
\langle 0, y_1 | H(y) | 0, y_2 \rangle
= \left( E_0(y_2) +\frac{1}{2}(y^{\mu}_2 -y^{\mu})^2 \right) 
\langle 0,y_1 |0,y_2 \rangle
+(y^{\mu}_2-y^{\mu})\langle 0,y_1 |
X_{\mu}-y_{2\mu}|0,y_2 \rangle.
\label{object1}
\end{align}
Because of (\ref{linear0}), the inner product 
$\langle 0,y_1 |0,y_2 \rangle$ satisfies
\begin{align}
\lim_{N\rightarrow \infty} 
(y^{\mu}_2-y^{\mu}_1) \langle 0,y_1 |0,y_2 \rangle
= 0.  \;\;\; (y_1, y_2 \in {\cal M})
\end{align}
This implies that the ground states at different points are orthogonal 
to each other in the large-$N$ limit. So we obtain
\begin{align}
\lim_{N\rightarrow \infty} 
K_N \langle 0,y_1 |0,y_2 \rangle
=  \delta^{(D)} (y_1-y_2). \;\;\; (y_1, y_2 \in {\cal M})
\label{orthogonality of coh st}
\end{align}
Here, we have introduced a proportionality constant $K_N$.
By applying (\ref{linear0}) and (\ref{orthogonality of coh st}) to 
(\ref{object1}), we obtain (\ref{lemma}).

Then, we prove (\ref{projectivity2prime}).
Note that $f(y + \epsilon_{\perp})$ can be written as
\begin{align}
f(y + \epsilon_{\perp})= 
\lim_{N\rightarrow \infty} 
\min_{| \alpha \rangle \in {\cal H}}
\langle \alpha | H(y+ \epsilon_{\perp})
| \alpha \rangle,
\label{fperp}
\end{align}
where $| \alpha \rangle$ shall be normalized 
as $\langle \alpha | \alpha \rangle =1$. 
Let us denote by ${\cal H}_{\rm c}$ the Hilbert space spanned by 
all the ground states in ${\cal M}$:
\begin{align}
{\cal H}_{\rm c} = {\rm span} \{|0,y' \rangle | y' \in {\cal M} \}.
\end{align}
The total space can then be decomposed as 
\begin{align}
{\cal H}= {\cal H}_{\rm c} \oplus \tilde{\cal H}.
\end{align}
By definition, $\tilde{\cal H}$ is the subspace on which 
the eigenvalues of Hamiltonian are always nonvanishing. 
Any element $|\beta \rangle $ in ${\cal H}_{\rm c}$ can be expanded as 
\begin{align}
| \beta \rangle =\sqrt{K_N}
\int d^Dy' \beta (y') |0, y' \rangle,
\label{coh st expansion}
\end{align}
where $\beta(y')$ is a function which is vanishing
unless $y'\in {\cal M}$.
Then a generic normalized vector $|\alpha \rangle $ takes the form,
\begin{align}
| \alpha  \rangle =  | \beta \rangle + | \tilde{\alpha}  \rangle,
\label{decomposition of alpha}
\end{align}
where $|\beta \rangle $ is given by (\ref{coh st expansion})
and $| \tilde{\alpha}  \rangle \in \tilde{\cal H}$.
The normalization condition is 
\begin{align}
\langle \beta |\beta \rangle +
\langle \tilde{\alpha} |\tilde{\alpha} \rangle =1.
\end{align}
By substituting (\ref{decomposition of alpha}) into 
(\ref{fperp}), we obtain 
\begin{align}
f(y + \epsilon_{\perp})= 
\lim_{N\rightarrow \infty} 
\min_{| \beta \rangle, | \tilde\alpha \rangle }
\left[
\langle \beta | H(y+ \epsilon_{\perp})
| \beta \rangle +
\langle \tilde{\alpha} | H(y+ \epsilon_{\perp})
| \beta \rangle 
\right.
\nonumber\\
+
\left.
\langle \beta | H(y+ \epsilon_{\perp})
| \tilde{\alpha} \rangle +
\langle \tilde{\alpha} | H(y+ \epsilon_{\perp})
| \tilde{\alpha} \rangle 
\right],
\label{fperp2}
\end{align}
For any $|\beta \rangle $ and $| \tilde{\alpha} \rangle $,
the second and the third terms are vanishing in the large-$N$ limit
because of (\ref{linear generalized}) and the orthogonality between 
$|\beta \rangle $ and $| \tilde{\alpha} \rangle $. 
Here, let us consider the case where $|\epsilon_{\perp}|$ is very small 
and put 
\begin{align}
\langle \beta |\beta \rangle 
= c|\epsilon_{\perp}|^a,
\end{align}
where $a$ and $c$ are non-negative constants.
Then, the last term in (\ref{fperp2}) is 
\begin{align}
(1-c|\epsilon_{\perp}|^a)\frac{\langle \tilde\alpha |
H(y+\epsilon_\perp)| \tilde\alpha \rangle}
{\langle \tilde\alpha | \tilde\alpha \rangle}.
\label{last term}
\end{align}
Note that from the definition of $\tilde{\cal H}$, 
the second factor is positive definite and is 
${\cal O}(|\epsilon_{\perp}|^0)$. 
Hence, (\ref{last term}) is ${\cal O}(|\epsilon_{\perp}|^0)$ unless 
$a=0$ and $c=1$.
Similarly, by using (\ref{lemma}), 
the first term in (\ref{fperp2}) can be estimated as 
${\cal O}(|\epsilon_{\perp}|^{a+2})$. 
Thus, we find that the minimum in (\ref{fperp2}) is saturated when
$a=0$ and $c=1$. (\ref{fperp2}) is now reduced to
\begin{align}
f(y + \epsilon_{\perp})= 
\lim_{N\rightarrow \infty} 
\min_{| \beta \rangle \in {\cal H}_{\rm c}}
\langle \beta | H(y+ \epsilon_{\perp})
| \beta \rangle. 
\label{fperp3}
\end{align}
Finally, by substituting the expansion (\ref{coh st expansion}) 
into (\ref{fperp3}) and using (\ref{lemma}), we obtain
\begin{align}
f(y + \epsilon_{\perp})= \frac{1}{2}
\int d^Dy' |\hat{\beta} (y')|^2
(y'^{\mu}-y^{\mu}-\epsilon^{\mu}_{\perp})^2.
\label{last eq}
\end{align}
Here, $\hat{\beta}(y')$ is the large-$N$ limit of 
the function which saturates the minimum in (\ref{fperp3}).
It satisfies $\int d^D y' |\hat{\beta}(y')|^2 =1$ and vanishes on the 
outside of ${\cal M}$.   
Obviously, the quantity on the right-hand side of 
(\ref{last eq}) is minimized when 
$\hat{\beta}(y')$ localizes at $y$. Thus, we finally 
obtain (\ref{projectivity2prime}).

\section{Properties of $W^{\mu \nu}$}
\label{Properties of W}
\subsection{Proof of (\ref{W tangent})}
Here, we show that $W^{\mu \nu }(y) $ is a tangent bivector, 
namely, it satisfies (\ref{W tangent}).
To this end, we use the following relation:
\begin{align}
\langle 0,y | [X^{\mu }, X^{\nu }] | 0,y \rangle 
\langle 0,y | X_{\nu} | n, y \rangle  
\sim  (E_{n}(y)-E_{0}(y))
\langle 0,y | X^{\mu} | n, y \rangle ,
\label{useful}
\end{align}
where $\sim $ stands for an equality for the leading-order terms in the 
large-$N$ limit.
The relation (\ref{useful}) is shown as follows. 
By using (\ref{condition}) and (\ref{linear0}), we obtain
\begin{align}
\langle 0,y | [X^{\mu }, X^{\nu }] | 0,y \rangle 
\langle 0,y | X_{\nu} | n, y \rangle  
& \sim 
\frac{1}{2} \langle 0,y |\{ X^{\nu}-y^{\nu}, [X^{\mu }, X_{\nu }-y_{\nu}]\} 
| n,y \rangle .
\label{interm}
\end{align} 
Then, note that 
$\{ X^{\nu}-y^{\nu}, [X^{\mu }, X_{\nu }-y_{\nu}]\} 
= 2[X^{\mu}, H(y)]$.
By substituting this into (\ref{interm})
we obtain (\ref{useful}).
Now, let us calculate 
\begin{align}
P^{\mu }{}_{ \rho }(y)W^{\rho \nu }(y) 
= \lim_{N \rightarrow \infty } 
\left[ 
2C_N {\rm Im} 
\left \{
 \sum_{n \neq 0} 
\frac{\langle 0, y | X^{\mu}|n,y \rangle 
\langle n, y | X^{\rho }| 0,y \rangle }
{E_n (y)- E_0 (y)} \langle 0, y | [X_{\rho}, X^{\nu}] |0,y \rangle
\right\}
\right] .
\end{align}
By using (\ref{useful}) and the completeness relation 
$\sum_{n=0}^{N-1} |n, y \rangle \langle n,y | =1 $, we obtain 
\begin{align}
P^{\mu }{}_{ \rho }(y)W^{\rho \nu }(y) = 
\lim_{N \rightarrow \infty } 
\left[ 
2C_N {\rm Im} 
\left \{ 
 \langle 0, y | X^{\mu }X^{\nu } | 0,y \rangle 
- \langle 0, y | X^{\mu }| 0,y \rangle 
\langle 0,y | X^{\nu } | 0,y \rangle 
\right\}
\right].
\end{align}
The second term is zero since $X^{\mu}$ are Hermitian, while the 
first term is equal to $W^{\mu \nu}(y)$.
Thus, we have shown (\ref{W tangent}).

\subsection{Proof of Jacobi identity}
Here, we show that the bracket defined in (\ref{P bracket})
satisfies the Jacobi identity. The Jacobi identity is equivalent to
\begin{align}
W^{\mu \nu}(y) \partial_{\mu} W^{\rho \sigma }(y)
+
W^{\mu \rho}(y) \partial_{\mu} W^{\sigma \nu }(y)
+
W^{\mu \sigma}(y) \partial_{\mu} W^{\nu \rho }(y)=0,
\label{Jacobi}
\end{align}
for $y \in {\cal M}$.

Let us consider an arbitrary polynomial $\Phi(X)$, where 
the coefficients and degree are assumed to be $N$-independent, 
and let us define a corresponding function by 
\begin{align}
\Phi (y) = \lim_{N \rightarrow \infty}
\langle 0,y | \Phi(X)| 0, y \rangle. 
\end{align}
Then, as shown below, the following relation holds: 
\begin{align}
\lim_{N \rightarrow \infty }
(-iC_N) \langle 0,y | [X^{\mu}, \Phi(X)]| 0,y \rangle 
= W^{\mu \nu} (y) \partial_{\nu }\Phi (y).
\label{general jacobi}
\end{align}
If we put $\Phi(X) = W^{\mu \nu}(X)
= -iC_N[X^{\mu}, X^{\nu}]+{\cal O}(1/C_N)$ in the above equation, 
(\ref{Jacobi}) immediately 
follows from the Jacobi identity of the matrix commutators.

Below, we show the relation (\ref{general jacobi}). 
The right-hand side of (\ref{general jacobi}) is the large-$N$ limit of 
\begin{align}
-iC_N \langle 0,y | [X^{\mu}, X^{\nu }] | 0, y \rangle 
\partial_{\nu } \langle 0,y | \Phi(X) | 0,y \rangle. 
\label{eq1 for Jacobi}
\end{align}
The derivative of $|0, y \rangle $ is obtained from the 
formula in the perturbation theory as
\begin{align}
\partial_{\mu}|0,y \rangle = 
\sum_{n\neq 0}
\frac{|  n,y \rangle \langle n,y| X_{\mu} | 0,y \rangle}
{E_n(y)-E_0(y) } + iA_{\mu}(y) |0,y \rangle,
\label{derivative of gs}
\end{align}
where $A_{\mu}(y)$ is the Berry connection. 
By substituting this, and using (\ref{useful}), we find that 
the large-$N$ limit of (\ref{eq1 for Jacobi}) 
is also equal to the left-hand side of (\ref{general jacobi}).
Thus, we obtain (\ref{general jacobi}).

\section{Construction of compatible structures}
\label{Construction of compatible structures}
In this appendix, we show that
on a symplectic manifold, if an arbitrary metric is given, 
one can find an almost complex structure and a Riemann
structure compatible with the 
symplectic form.

We assume that we are given a metric $g$ on a symplectic
manifold $({\cal M }, \omega ) $.
For any tangent vector $u \in T{\cal M}$, let us consider 
two mappings defined by
\begin{align}
&u \mapsto \omega(u,\cdot ) \in T^*{\cal M},
\nonumber\\
&u \mapsto g(u,\cdot ) \in T^*{\cal M}.
\end{align}
Since both of $g$ and $\omega$ are nondegenerate, 
these mappings define isomorphisms $T{\cal M}\rightarrow T^*{\cal M}$.
Then we can define an isomorphism $A: T{\cal M}\rightarrow T{\cal M}$
by 
\begin{align}
\omega(u,v) = g(Au,v). \;\;\; (u,v\in T{\cal M}) 
\end{align}
The isomorphism $A$ is antisymmetric, since
\begin{align}
g(A^Tu, v)=g(u,Av)=g(Av,u)
=\omega (v,u) = -\omega(u,v)=-g(Au,v).
\end{align}
The combination $AA^T$ is symmetric. By 
diagonalizing it, we can write
\begin{align}
AA^T = B\
{\rm diag}(\lambda_1, \lambda_2, \cdots, \lambda_{{\rm dim }{\cal M}})
 B^{-1},
\label{AAT}
\end{align}
where $\lambda_i >0 \; (i=1,2,\cdots,{\rm dim}{\cal M})$. 

An almost complex structure $J$ on ${\cal M}$ can be defined 
by a polar decomposition of $A$ as
\begin{align}
J= (AA^T)^{-1/2}A,
\label{general def of J}
\end{align}
where $(AA^T)^a$ is defined by replacing
$\lambda_i$ with $\lambda_i^a$ 
in the right-hand side of (\ref{AAT}).
Obviously, $J$ is antisymmetric and satisfies 
$J^2= -1$.

In general, an almost complex structure is said to be compatible 
with the symplectic structure if 
for any $u,v \in T{\cal M}$ it satisfies
\begin{align}
\omega(Ju, Jv)=\omega(u,v),
\end{align}
and 
\begin{align}
\omega(u, Ju)>0.
\label{2nd compatibility}
\end{align}
It is easy to see that the definition (\ref{general def of J})
satisfies the both conditions and thus it gives a 
compatible almost complex structure.
In fact, one can check the first condition by
\begin{align}
\omega(Ju,Jv)=g(AJu,Jv)=g(JAu,Jv)=g(Au,J^TJv)
=g(Au,v)=\omega(u,v),
\end{align}
and the second condition by
\begin{align}
\omega(u,Ju)=g(Au,Ju) = -g(JAu,u)
= g((AA^T)^{1/2}u,u) 
= g((AA^T)^{1/4}u,(AA^T)^{1/4}u)
>0.
\end{align}

Once a compatible pair of the symplectic form and 
almost complex structure $(\omega, J)$, one can define
a new compatible metric $\tilde{g}$ by 
\begin{align}
\tilde{g}(u,v)= \omega(u,Jv).
\end{align}
The positivity is guaranteed by 
(\ref{2nd compatibility}).

\section{Explicit forms of $J$ and $\tilde{g}$}
\label{Explicit form of J and g}
In this appendix, we derive 
(\ref{ex form of J})
and (\ref{ex form of gtilde}).

Because of (\ref{W tangent}),
the complex structure $J$ and the metric $\tilde{g}$ can be written 
for $y \in {\cal M}$ as 
\begin{align}
&J(y)= 
\frac{1}{\sqrt{W(y)W^T(y)}}
W(y)P(y),
\nonumber\\
&\tilde{g}(y)= 
\frac{1}{\sqrt{W(y)W^T(y)}}P(y).
\label{J tildeg}
\end{align}
Here, let us consider the equation 
(\ref{useful}).
This shows that the multiplication of $W$ on 
$\langle 0,y | X^{\nu}| n,y \rangle $ produces 
the factor $-iC_N(E_n(y)-E_0(y))$.
This also implies that the multiplication of 
$\frac{1}{\sqrt{WW^T}}$ yields $\frac{1}{C_N(E_n(y)-E_0(y))}$.
Then, applying these calculation to (\ref{J tildeg}),
where $P$ is written as (\ref{projection}),
we obtain (\ref{ex form of J})
and (\ref{ex form of gtilde}).

Note that (\ref{useful}) implies that $E_n-E_0 = {\cal O}(1/C_N)$ or 
otherwise $\langle 0,y | X^{\mu}|n,y \rangle =0$. 
So the above multiplication is well-defined in the large-$N$ limit.

\section{Geometric quantization on K\"{a}hler manifold}
\label{Geometric quantization on Kahler manifold}
In this appendix, we briefly review the geometric quantization on a 
compact K\"{a}hler manifold \cite{gqtext,gqreview}. 
The geometric quantization is a generalization of the process of 
up-lifting a classical mechanics to the corresponding quantum mechanics.
In the usual setup of Hamilton mechanics, 
the classical phase space is an even dimensional flat space. The coordinates 
$(p_i, q_i)$ and observables $f_n(p,q)$ are promoted to 
some Hermitian operators in going from the classical mechanics 
to the quantum mechanics.
The geometric quantization is a generalization of this problem
such that the phase space at the starting point is not 
a flat space but a non-trivial symplectic manifold.
The goal of the geometric quantization is to construct 
a quantum Hilbert space and quantum mechanical operators 
represented on that space, for given 
symplectic manifold ${\cal M}$ and 
classical observables on ${\cal M}$.

We consider a K\"{a}hler manifold, which has a compatible triple 
$(\omega , J, \tilde{g})$, where $\omega , J$ and $ \tilde{g}$ are 
symplectic, complex and Riemann structures, respectively.
For simplicity, let $\omega$ be normalized 
in such a way that $\omega /2\pi$ represents 
an integral cohomology class and its integral over every closed two-cycle in 
${\cal M}$ is equal to 1.
At least locally (i.e. on each coordinate patch), one can introduce the symplectic potential as
\begin{align}
\omega = d\theta.
\end{align}
On an overlap of two patches, two coordinates are related by 
a symplectomorphism preserving $\omega $. Under the 
symplectomorphism, $\theta $ transforms like a $U(1)$ gauge field.
If ${\cal M}$ possesses non-trivial one-cycles, there will be an ambiguity 
in the definition of $\theta$ coming from the choice of the holonomy. 
For each choice, the following construction works. 
So we take one of them and fix it in the following discussion. 

We first introduce the notion of the prequantum line bundle. 
In the case of quantum mechanics (i.e. when ${\cal M}$ is flat,)
wave functions usually depend only on $q_i$, or in the 
momentum representation, only on $p_i$. 
The sections of the prequantum line bundle correspond to 
more general functions of the form $\psi (p_i,q_i)$ 
and form a larger reducible Hilbert space. 
In order to obtain a relevant Hilbert space, one needs a 
truncation to the irreducible space.
This procedure (choosing a polarization) is
discussed later. 
To define the prequantum line bundle,
let us introduce covariant derivatives as
\begin{align}
D_{a}= \partial_a -i A_{a}.
\end{align}
Here the new gauge field $A$ is defined by 
\begin{align}
A = n \theta,
\end{align}
where $n$ is an arbitrary fixed integer. 
We denote by $F$ the curvature of $A$. 
The integer $n$ then corresponds to the monopole charge of $F$.
The prequantum line bundle is just the associated 
complex line bundle with the curvature given by $F$.
Let $\Phi $ be a section of the line bundle. 
It transforms as $\Phi \rightarrow e^{i\Lambda}\Phi$
when the gauge field transforms as $A \rightarrow A+d\Lambda$.
The inner products for sections are defined by using the Liouville measure 
of $\omega$. 

Next, we introduce prequantum operators.
For a given function $f$ on the K\"{a}hler manifold,
the prequantum operator is defined by
\begin{align}
{\cal P}(f)=f -i \xi_f^a D_a,
\end{align}
where $\xi_f^a$ is the Hamilton vector field of $f$ defined by
\begin{align}
i_{\xi_f}\omega = -d f.
\end{align}
The prequantum operators act on the space of sections introduced above.
One of the most important properties of those operators is
\begin{align}
[{\cal P}(f), {\cal P}(g)]= i{\cal P}(\{f,g \}),
\end{align}
where, $\{\; , \; \}$ is the Poisson bracket defined by
\begin{align}
\{f,g \} = i_{\xi_f}i_{\xi_g}\omega.
\end{align}
Namely, the Poisson bracket is mapped to the commutator under 
the action of ${\cal P}$.

In order to obtain a relevant irreducible Hilbert space, one needs to choose 
a polarization of the wave functions. 
On K\"{a}hler manifolds, one can take the so-called holomorphic
polarization. By using the complex structure, we 
can define holomorphic and anti-holomorphic covariant derivatives,
\begin{align}
D_{z_i}=\partial_{z_i}-iA_{z_i}, \;\;\;
D_{\bar{z}_i}=\partial_{\bar{z}_i}-iA_{\bar{z}_i},
\end{align}
where $i$ runs from $1$ to the complex dimension of the manifold.
The polarization condition is then written as 
\begin{align}
D_{\bar{z}_i} \Phi =0,
\label{polarization}
\end{align}
for any $i$, where $\Phi $ is a section of the prequantum line bundle.
Square integrable holomorphic sections satisfying 
(\ref{polarization})
form a subspace of the prequantum Hilbert space and this defines 
the true quantum Hilbert space. 
The quantum Hilbert space generally forms an irreducible representation 
space of the Poisson algebra.

The dimension of the quantum Hilbert space is finite in general 
for the K\"{a}hler polarization on a compact K\"{a}hler manifold. 
Through an index theorem, the dimension is given by an integral
over ${\cal M}$ of a wedge product of the Todd class 
on ${\cal M}$ and the Chern character of $F$. 
For example, a Riemann surface with genus $g$, the dimension is 
given by $n+1-g$. 
For general cases, in the large-$n$ limit, 
the dimension is approximately given by $n^{{\rm dim}_C{\cal M}}$,
where ${\rm dim}_C{\cal M}$ is the complex dimension of ${\cal M}$.

If a prequantum operator is closed on the polarized wave functions, 
it can be promoted to an operator on the polarized Hilbert space.
If this is not the case, however, things become a little complicated.
For such cases, 
prescriptions to construct operators in the polarized space  
are known for some particular cases (see \cite{gqtext,gqreview} and 
references therein.),
though we will not treat those cases in this paper.


\end{document}